\documentclass[%
 aip,
 amsmath,amssymb,
 reprint,%
]{revtex4-2}

\usepackage{graphicx}
\usepackage{dcolumn}
\usepackage{bm}
\usepackage{physics}
\usepackage[utf8]{inputenc}
\usepackage[T1]{fontenc}
\usepackage{mathptmx}
\usepackage{etoolbox}

\makeatletter
\def\@email#1#2{%
 \endgroup
 \patchcmd{\titleblock@produce}
  {\frontmatter@RRAPformat}
  {\frontmatter@RRAPformat{\produce@RRAP{*#1\href{mailto:#2}{#2}}}\frontmatter@RRAPformat}
  {}{}
}%
\makeatother

\begin{document}

\preprint{AIP/123-QED}

\title{Nonlinear Equation for Dust Drift Waves}
\author{H. Saleem}
\thanks{saleemhpk@hotmail.com}
\affiliation{Theoretical Research Institute, Pakistan Academy of Sciences (TRIPAS), 3-Constitution Avenue, G-5/2, Islamabad (44000), Pakistan.}
\affiliation{School of Natural Sciences (SNS), National University of Sciences and Technology (NUST), H-12, Islamabad (44000), Pakistan.}
\affiliation{Department of Scpace Science, Institute of Space Technology (IST), 1-Islamabad Expressway, Islamabad (44000), Pakistan.}
\date{Jan. 18, 2026}

\begin{abstract}
A nonlinear equation for dust drift waves is derived assuming two-dimensional propagation and ignoring the role of dust acoustic waves. Both the nonlinear dust density term and the dust vorticity term are taken into account. If vorticity term is ignored, the equation gives two dimensional solitary waves and if density nonlinearity effect is discarded, then it reduces to Hasegawa-Mima equation for dust drift waves which admits dipole vortex solutions.

\end{abstract}

\bigskip 
\maketitle
\section{Introduction}
Dusty plasmas can support a variety of very low-frequency waves that give rise to different nonlinear structures, including solitons, vortices, and shock waves \cite{shukla2003solitons}. Such plasmas are commonly found in the magnetospheres of Jupiter and Saturn, as well as in cometary environments, where massive charged dust grains behave as an additional plasma fluid and support new low-frequency oscillations and wave modes. Dusty plasmas are also present in the interstellar medium (ISM). The presence of charged dust significantly modifies the natural modes of an electron ion plasma. While static dust alters the characteristics of conventional plasma waves, the inclusion of dust-fluid dynamics leads to the emergence of several new wave modes and collective phenomena \cite{verheest2012waves, shukla2015introduction}.

An inhomogeneous magnetized electron--ion plasma supports a low-frequency electrostatic mode known as the drift wave, which plays a fundamental role in cross-field particle and energy transport \cite{horton1999drift, weiland1999collective}. The existence of electrostatic drift waves was first predicted theoretically by Rudakov and Sagdeev \cite{rudakov1960oscillations} and Kadomtsev and Pogutse \cite{kadomtsev1962drift}, and was subsequently confirmed experimentally \cite{d1963low}. Since both drift waves and ion-acoustic waves (IAWs) are low-frequency electrostatic modes, they are generally coupled in plasma systems \cite{kadomtsev1962drift}. Consequently, extensive investigations have been carried out on IAWs as well as on their coupled dynamics with drift waves. However, Hasegawa and Mima \cite{hasegawa1978pseudo} were the first to examine pure drift waves by neglecting ion parallel motion, thereby eliminating the coupling with IAWs. Their model admits dipolar vortex solutions. In contrast, pure ion-acoustic waves can support a wide range of nonlinear structures, including solitons, shocks, double layers, and vortices. The possibility of solitary structures associated with drift waves was first investigated by Nozaki \textit{et al.} \cite{nozaki1974propagation} and later by Gell  \cite{gell1977drift}; however, in both studies the drift waves remained coupled to IAWs. Following these pioneering works, most nonlinear studies of drift waves considered their coupling with ion-acoustic modes in various plasma configurations. Recently, the nonlinear dynamics of pure drift waves, uncoupled from IAWs, have been investigated in 2024 \cite{Saleem2024}. By neglecting ion parallel motion, the Korteweg--de Vries (KdV) equation has been derived for drift waves, completely excluding ion-acoustic effects. It was shown that drift-wave solitons correspond to density depressions, in contrast to ion-acoustic solitons, which are associated with density enhancements. This behavior arises because, in the absence of ion parallel motion, the ions cannot effectively counterbalance the nonlinear electron density response, which contributes a negative nonlinear term in the KdV equation.

More recently \cite{saleem2026exploring}, a new nonlinear equation describing electrostatic drift waves in electron–ion plasmas has been derived, which admits both soliton and vortex solutions in different limiting cases. It  demonstrates that a general nonlinear drift wave equation must incorporate both the scalar electron-density nonlinearity, which naturally arises from the Boltzmann electron response, and the vector nonlinearity associated with the ion polarization drift. Neglecting the electron-density nonlinearity reduces the equation to the well-known Hasegawa–Mima equation \cite{hasegawa1978pseudo}. Motivated by these findings, it is important to investigate the nonlinear dynamics of pure dust drift waves by including the nonlinear dust density contribution alongside the polarization drift nonlinearity, following an approach similar to that adopted in Ref. \cite{saleem2026exploring}.


\section{Basic equations}
We consider a plasma consisting of electrons, ions and heavier dust particles embedded in a constant magnetic field ${\bf B}_0=B_0 \hat{z}$. It is assumed that the density is inhomogenoeus along x-axis such that $\nabla n_{j0}=- \hat{x} \mid  \frac{dn_{j0}}{dx} \mid $ where $j=(e,i,d)$ and $\kappa_j = \mid \frac{1}{n_{j0}} \frac{d n_{j0}}{dx} \mid $ is constant. The dust drift wave (DDW) is a very low frequency wave with frequency $\omega$ much less than the dust gyro frequency $\Omega_i = \frac{e Z_d  B_0}{m_d c}$ and here $Z_d$ denotes positive or negative charge on a dust particle which is assumed to be constant for simplicity whereas $m_d$ is the mass of dust particle. Since $\omega << \Omega_d$, therefore electrons and ions are assumed to be thermal obeying Boltzmann density distribution in the electrostatic field of the wave ${\bf E} = - \nabla \varphi$ as given below,   \\
\begin{equation}
n_e = n_{e0} e^{ e \varphi/ T_e } 
\end{equation}
and, 
\begin{equation}
n_i = n_{i0} e^{- e \varphi/ T_i } 
\end{equation}
where $T_e(T_i)$ are constant electron (ion) temperatures.
Momentum conservation equation for cold ideal dust fluid is, 
\begin{equation}
m_d n_d (\partial_t + {\bf v}_d \cdot \nabla) {\bf v}_d = q_d n_d ({\bf E}+ \frac{1}{c} {\bf v}_d \times {\bf B}_0)
\end{equation}
where $q_d=e Z_d$.
Taking cross product with unit vector $\hat{z}$, we obtain the perpendicular part of dust velocity,
\begin{equation}
{\bf v}_{d \perp} \simeq - \frac{c}{B}_0 \nabla_{\perp} \varphi  \times \hat{z}- \frac{c}{B_0 \Omega_d} (\partial_t + {\bf v}_d \cdot \nabla) \nabla_{\perp} \varphi = {\bf v}_E+ {\bf v}_{Pd}
\end{equation}
where ${\bf v}_E$ and ${\bf v}_{Pd}$ are the electric and polarization drifts, respectively.

It is important to note that most early linear and nonlinear investigations of conventional drift waves focused on the coupled dynamics of ion acoustic waves (IAWs) and drift waves \cite{nozaki1974propagation, gell1977drift, goswami1977finite}. This approach was subsequently extended to more complex plasma environments, including bi-ion \cite{shan2019double, shan2020kappa} and dusty plasmas \cite{shukla1991linear, saleem2004rotation}, where similar theoretical frameworks were employed. In contrast, \citeauthor{hasegawa1978pseudo} \cite{hasegawa1978pseudo} derived a nonlinear equation for two-dimensional drift waves by neglecting ion parallel motion. The resulting Hasegawa--Mima equation treats the linear drift-wave dynamics and the nonlinear ion polarization drift on an equal footing and has become a cornerstone of drift-wave theory. However, it has subsequently been recognized that a complete nonlinear description of drift waves should also include the nonlinear electron density contribution, which arises naturally from the Boltzmann electron response. The recently derived nonlinear equation for conventional drift waves \cite{saleem2026exploring} incorporates both essential nonlinearities, namely the ion polarization drift nonlinearity and the nonlinear electron density term, thereby providing a more comprehensive description of drift-wave dynamics.

To investigate the one dimensional $1D$ and two dimensional (2D) nonlinear dynamics of pure dust drift waves, we ignore the parallel component of wave vector assuming $k_z<< k_{\perp}$. Our aim is to derive a nonlinear equation applicable to both positively and negatively charged dust fluids. The mass conservation equation for cold dust fluid is,
\begin{equation}
\partial_t n_d + \nabla \cdot (n_d {\bf v}_d)=0
\end{equation}
Equation (4) yields,
\begin{equation}
\nabla \cdot {\bf v}_{d \perp} \simeq - \frac{c}{B_0 \Omega_d } (\partial_t + {\bf v}_E \cdot \nabla) \nabla_{\perp}^2 \varphi
\end{equation}
The continuity equation can be expressed as follows,
\begin{equation}
(\partial_t + {\bf v}_E \cdot \nabla) \frac{\tilde {n}_d}{n_{d0}} + \frac{\nabla n_{d0}}{n_{d0}} \cdot {\bf v}_E 
- \frac{c}{B_0 \Omega_d } (\partial_t + {\bf v}_E \cdot \nabla) \nabla_{\perp}^2 \varphi = 0
\end{equation}
where $n_d = n_{d0} + \tilde {n}_d$. Assuming $e \varphi/ T_e << 1$ and $e \varphi/ T_i << 1$ along with quasi-neutrality,
\begin{equation}
n_i - n_e + q_{\alpha} n_d =0
\end{equation}
we write,
\begin{equation}
n_d = \frac{1}{Z_d} \left[ n_{e0} (1+ e \varphi/ T_e + \frac{1}{2} (e \varphi/ T_e)^{1/2} + ...) \right]  
\end{equation}

$$
 - \frac{1}{Z_d} \left[ n_{i0} (1 - \sigma e \varphi/ T_e + \frac{1}{2} \sigma^2 (e \varphi/ T_e)^{1/2} + ...) \right]
$$
where $\sigma= \frac{T_e}{T_i}$. It gives,
\begin{equation}
\frac{\tilde {n}_d}{n_{e0}} \simeq \frac{1}{Z_d} \left[ (\frac{n_{e0}+ \sigma n_{i0}}{n_{d0}} ) (e \varphi/ T_e) +\frac{1}{2} (\frac{n_{e0} - \sigma^2 n_{i0}}{n_{d0}} ) (e \varphi/ T_e)^2  \right] 
\end{equation}
The higher order terms have been ignored. To obtain linear dispersion relation, 
let us consider propagation of the wave along y-axis. Then in linear limit $\varphi \propto e^{\iota (k_y y - \omega t)}$ and in Fourier space dust continuity equation reduces to,
\begin{equation}
\partial_t \frac{\tilde {n}_d}{n_{d0}} + \frac{\nabla n_{d0}}{n_{d0}} \cdot {\bf v}_E - \frac{c}{B_0 \Omega_d} \partial_t \nabla_{\perp}^2 \varphi =0
\end{equation}
It yields the linear dispersion relation in the following standard form,
\begin{equation}
\omega = \frac{\omega_d^{\ast}}{1 + \rho_{sd}^2 k_y^2}
\end{equation}
where $\omega_d^{\ast} = {\bf v}_{Dd} \cdot {\bf k}=v_{Dd}k_y$, ${\bf v}_d = D_d \nabla \ln n_{d0} \times \hat{z}=(D_d \kappa_d) \hat{y}$, $D_d=\frac{c T_{eff}}{e Z_d B_0}$, $\rho_{sd}=\frac{c_{sd}}{\Omega_d}$, $c_{sd}=(\frac {T_{eff}}{m_d})^{1/2}$ and $T_{eff}= \frac{Z_d^2 T_{i0} T_{e0} n_{d0}}{T_{i0} n_{e0}+ T_{e0} n_{i0}}$. We notice that $\omega_d^{\ast}<0$ for $Z_d<0$ and vice a versa because ${\bf v}_d$ depends upon the sign of $Z_d$.


\section{Derivation of Nonlinear Equation}
To write the dust continuity equation in normalized form, we define $t^{\prime} = \frac{c_{sd}}{L_d} t$, $\nabla^{\prime}= \rho_{sd} \nabla$ and $\Phi = \frac{e Z_d \varphi}{T_{eff}}$ where $L_d=\frac{1}{\kappa_d}$. Then perturbed dust density can be expressed as,
\begin{equation}
\frac{\tilde {n}_d}{n_{d0}} \simeq (\Phi + \frac {\alpha}{2} \Phi^2)
\end{equation}
where higher order terms have been ignored and $\alpha =  (\frac{T_{i0}^2 n_{e0} - T_{e0}^2 n_{i0}}{n_{d0} T_{i0}^2}) \frac{T_{eff}^2}{Z_d^3 T_{e0}^2} >0$. Since $n_{e0} << n_{i0}$ in case of negative dust, therefore $\alpha$ is generally positive for both positive and negative dust particles. But the sign of $\rho_{sd}$ and hence of $\nabla$ depends upon the sign of $Z_d$.
Dropping the superscript prime $(\prime)$ from all quantities, the nonlinear equation can be written in normalized form as follows,
\begin{equation}
\partial_t \left[(\Phi - \nabla _{\perp}^2 \Phi) + \frac{\alpha}{2} \Phi^2 \right] + (\nabla_{\perp} \Phi \times \hat{z} \cdot \hat{x}) 
\end{equation}
$$
+ \frac{L_d}{\rho_{sd}} (\nabla_{\perp}\Phi \times \hat{z} \cdot \nabla_{\perp} ) \nabla_{\perp}^2 \Phi = 0
$$
This is a general nonlinear equation for dust drift waves which is very similar to the form of electron ion plasma case discussed in Ref. \cite{,saleem2026exploring} except that this equation depends upon the dust mass and charge. Therefore the temporal and spatial scales are very different.

\section{Limiting Cases}
Now we look into the different cases of nonlinear dust drift wave equation (14).
\subsection{!D and 2D Solitons}
If convective derivative in polarization drift is ignored, then the vorticity term disappears and Eq. (14) reduces to,
\begin{equation}
\partial_t \left[(\Phi - \nabla _{\perp}^2 \Phi) + \frac{\alpha}{2} \Phi^2 \right] + (\nabla_{\perp} \Phi \times \hat{z} \cdot \hat{x}) =0
\end{equation}
In one-dimensional case $\nabla = \hat{y} \partial_y$, the above equation becomes,
\begin{equation}
\partial_t \left[(\Phi - \partial_y^2 \Phi) + \frac{\alpha}{2} \Phi^2 \right] + (\nabla_{\perp} \Phi \times \hat{z} \cdot \hat{x}) =0
\end{equation}
Solutions of Eqs. (15) and (16) have been discussed in Refs. \cite{saleem2026exploring} and \cite{Saleem2024}, respectively, for the case of electron ion plasma.
\subsection{Hasegawa-Mima Like Equation}
If density nonlinearity is neglected by assuming $\Phi^2 << \Phi$, then Eq. (14) reduces to Hasegawa-Mima equation for dust drift waves \cite{shukla1991linear}. In this limit, the Eq. (14) takes the form, 
\begin{equation}
\partial_t \left[(\Phi - \nabla _{\perp}^2 \Phi)  \right] + (\nabla_{\perp} \Phi \times \hat{z} \cdot \hat{x}) 
\end{equation}
$$
+ \frac{L_d}{\rho_{sd}} (\nabla_{\perp}\Phi \times \hat{z} \cdot \nabla_{\perp} ) \nabla_{\perp}^2 \Phi = 0
$$
Solution of this equation has been presented in Ref. \cite{saleem2026exploring} for usual drift waves in electron ion plasma.

\subsection{Electron Ion Plasma}
If positive dust is considered for $Z_d=1$ then $\Omega_d >0$ and by using $n_{i0}=0$ and $n_{d0}=n_{e0}$, we get $T_{eff} = T_e$ and $\alpha =1$, In this limit $\rho_{sd}$ is the positive dust gyro radius at electron temperature and $L_d$ is the density gradient scale length. Then ions in this case are the heavier dust ions. Therefore, Eq. (14) becomes exactly the same as Eq. (13) of Ref. \cite{saleem2026exploring}.

\subsection{Negative Dust}
For the case of negative dust $Z_d<0$, we have  $\Phi = - \Phi$ and hence $\alpha$ remains positive. The dust Larmor radius  changes sign with the sign of $Z_d$. The Eq. (14) for negative dust becomes,
\begin{equation}
\partial_t \left[(\Phi - \nabla _{\perp}^2 \Phi) - \frac{\alpha}{2} \Phi^2 \right] - (\nabla_{\perp} \Phi \times \hat{z} \cdot \hat{x}) 
\end{equation}
$$
+ \frac{L_d}{\rho_{sd}} (\nabla_{\perp}\Phi \times \hat{z} \cdot \nabla_{\perp} ) \nabla_{\perp}^2 \Phi = 0
$$
This equation is different from Eq. (14) because signs of two terms become negative.
\vspace{.3 cm}

\section{Summary}
A nonlinear equation for dust drift waves has been derived taking into account the role of vorticity along with the nonlinear density term which appears due to the Boltzmann distributions of electrons and ions. The form of this equation for the case of positive dust is very similar to the case of electron ion plasma presented in a recent article \cite{saleem2026exploring}. However, the characteristic spatial and temporal scales differ significantly because the dynamics are controlled by the much heavier dust fluid, whose response depends on the dust mass and charge.

If the contribution of dust vorticity is ignored the equation reduces to the form similar to two dimensional Korteweg de-Vries (KdV) equation (15) and it admits solitary structure solution in 2D. If propagation of waves is considered only perpendicular to the density gradient and ambient magnetic field i.e. along y-axis, then Eq. (15) reduces to the standard form of KdV equation which gives 1D solitons. On the other hand, if density nonlinear term which is proportional to $\Phi^2$ is ignored, the Eq. (14) takes the form of Hasegawa-Mima equation \cite{hasegawa1978pseudo} and it is  solved using cylinrical coordinates \cite{larichev1976two}. Nevertheless, for drift waves the polarization drift is generally much smaller than the electric drift, suggesting that density nonlinearity is likely the dominant nonlinear mechanism.

The general formulation has also been extended to plasmas containing negatively charged dust. In this case, the structure of the governing equation is modified, with two nonlinear terms changing sign. The equations derived for both positively and negatively charged dust can be applied to the study of drift-wave phenomena in a variety of astrophysical and space dusty plasmas, including the magnetospheres of Jupiter and Saturn, the interstellar medium (ISM), and cometary tails.

\bigskip

\section{Data Availability Statement}
The data used for the preparation of the presented results is included in the text of this work.

\bibliography{bibliography}

\end{document}